\documentclass{epl}
\def \reo {$\mathrm{(TMTSF)_2ReO_4}$\,}
\def \areo {$\mathrm{ReO_4^-}$\,}
\def \clo {$\mathrm{(TMTSF)_2ClO_4}$\,}
\def \pf {$\mathrm{(TMTSF)_2PF_6}$\,}
\def \tmtsfx {$\mathrm{(TMTSF)_2X}$\,}
\def \tmtsf {$\mathrm{TMTSF}$\,}
\def \tao {$\mathrm{T_{AO,2}(P)}$\,}
\def \taot {$\mathrm{T_{AO,3}(P)}$\,}

\title{Self-organization of charge under pressure in the organic conductor \reo}
\author{C.Colin\inst{1}\footnote{Present address : Department of Chemical Physics, Materials Science Centre, Nijenborgh 4, 9747 AG Groningen, The Netherlands} \and P.Auban-Senzier\inst{1} \and C.R.Pasquier\inst{1} \and K.Bechgaard \inst{2}}

\institute{                    
  \inst{1} Laboratoire de Physique des Solides, UMR8502, Bat 510, Centre Universitaire, F-91405 ORSAY C\'edex, France \\
  \inst{2} Department of Chemistry, H.C.Oersted Institute, Univeristetsparken 5, DK2100, Copenhagen, Denmark
}

\begin{document}
\maketitle

\begin{abstract}
\reo presents a phase coexistence between two anion orderings defined by their wave vectors $q_2=(1/2,1/2,1/2)$ and $q_3=(0,1/2,1/2)$ in a wide range of pressure (8-11kbar) and temperature. From the determination of the anisotropy of the conductivity and the superconducting transitions in this regime we were able to extract the texture which results from a self-organization of the orientations of the \areo anions in the sample. At the lowest pressures, the metallic parts, related to the q$_3$ order, form droplets elongated along the a-axis embedded in the semiconducting matrix associated with the q$_2$ order. Above 10kbar, filaments along the a-axis extend from one end of the sample to the other nearly up to the end of the coexistence regime. A mapping of the system into an anisotropic Ising lattice is satisfactory to analyze the data. 
\end{abstract}

Self-organization of electronic charge in strongly correlated electron systems is now commonly observed in many families of materials such as cuprates\cite{Tranq95, Ando02} or manganites \cite{Dagotto01} but the precise texture and the scale, at which phase separation occurs, is still the subject of a large controversy.
In organic charge transfer salts, the question of phase coexistence has been only recently addressed. Clear experimental evidence of the texture is still lacking in these materials but electronic phase coexistence between antiferromagnetism and superconductivity has been clearly established in the two dimensional organic conductor $\kappa$-(BEDT-TTF)$_2$Cu[N(CN)$_2$]Cl under hydrostatic pressure \cite{Lefebvre00}. On the other hand, the possibility of charge-ordered stripes in the 2D compounds $\alpha$-(BEDT-TTF)$_2$I$_3$ \cite{Takano01} or $\theta$-(BEDT-TTF)$_2$RbZn(SCN)$_4$ \cite {Miyagawa00} has been theoretically proposed \cite{Seo00} following NMR experiments performed in these compounds.  The most studied charge transfer salts belong to the \tmtsfx family\cite{Jerome05}. These quasi-one dimensional compounds present a natural strong anisotropy of conduction along the three crystallographic axes \textit{a}, \textit{b'} and \textit{c*} with typical ratios of the resistivities $1:10^{-2}:10^{-5}$ respectively. At low temperatures, the metallic state is replaced by either a spin density wave or a superconducting state as a function of the applied pressure. In \pf, these two instabilities coexist in a 0.8kbar range of pressure and the existence of a macroscopic phase coexistence has been highlighted\cite {Vuletic02}. In this regime and in presence of an external magnetic field, the upper critical field was shown to diverge which was understood as a direct proof of a lamellar structure at the macroscopic level\cite{Lee02}. In the \tmtsfx family, the presence of non centro-symmetrical anions modifies this simple picture by introducing additional degrees of freedom. In this respect, \reo presents specific properties: at ambient pressure\cite{Parkin81}, at T$_{AO,2}$(P=1bar)=179 K, a weakly first order phase transition occurs from a metallic state where the tetrahedral anions are disordered toward a semiconducting state characterized by the so-called wave vector $q_2=(1/2,1/2,1/2)$ \cite{Moret82}. As hydrostatic pressure is increased, the semiconducting state is stabilized at lower temperatures. Above 8kbar, a new anion ordering transition occurs at \taot which is second order and increases with pressure with \taot$>$\tao. This metallic ordered state is characterized by the so-called wave vector $q_3=(0,1/2,1/2)$ \cite{Moret86}. X-rays measurements also established the coexistence of the two anion orderings below \tao from 8 to 11 kbar typically in a wide range of temperature.

The present work identifies the texture of the sample in the regime of pressure and temperature where both anion orderings coexist in \reo near the observed end point of the first order phase transition line. The texture is a direct consequence of the self-organization of the orientation of the \areo anions in the coexistence regime which manifests itself in the electrical properties. We have performed conductivity measurements along the different axes which clearly establish an exceptional increase of the anisotropy of the conduction in this regime by the formation of metallic filaments or droplets elongated along the \tmtsf chains (\textit{a}-axis).

We have performed four points resistivity measurements on three different samples of sizes $2.5\times0.125\times0.05$ mm$^{3}$, $1.5\times0.5\times0.075$ mm$^{3}$ and  $2.6\times0.5\times0.06$ mm$^{3}$ respectively. The resistivity, $\rho_a$, along the \textit{a} crystallographic axis was measured on the first sample. Similarly, $\rho_b$ and $\rho_c$ were measured independently on the two last samples with the current applied along the \textit{b'} and \textit{c*} crystallographic axes respectively. For resistances up to 100k$\Omega$, a standard low frequency ac lock-in technique was used with an applied current of 1$\mu$A. For larger resistances, a dc measurement was used with limited currents to avoid heating or non-linear effects which may arise at large currents. This was checked by decreasing the current by a factor of 5 or 10 with no change in the obtained resistance. The lowest applied current was 0.5nA which gives a maximum measured resistance of about 2G$\Omega$. Pressure was applied in a CuBe clamp using silicon oil as the transmitting medium. The pressure is deduced from the resistance of a manganine gauge at ambient temperature. The temperature was carefully measured in the pressure cell down to 15K using a copper-constantan thermocouple. At lower temperatures, we have used a RuO$_2$ thermometer glued on the outside of the pressure cell. As samples are measured in different runs, the \tao and \taot lines were used to readjust pressure from one sample to the other. 
\begin{figure}
\centerline{\includegraphics[width=0.85\hsize]{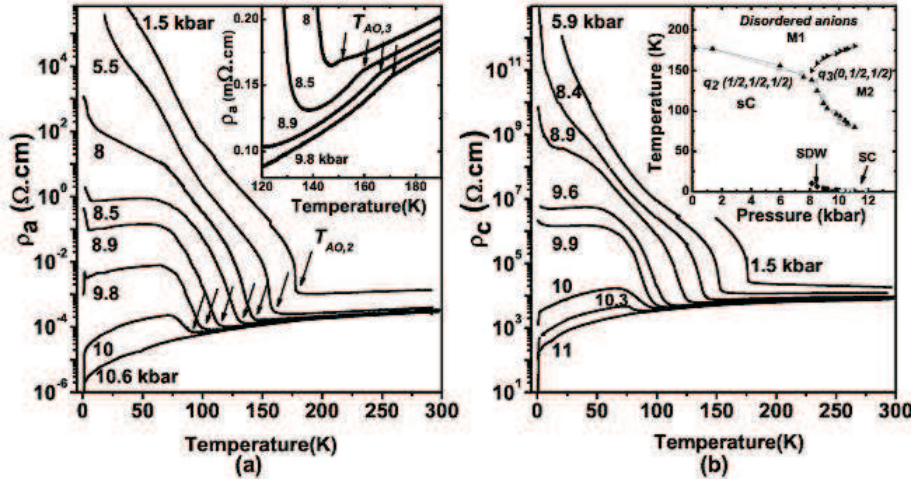}}
\caption{(a) Temperature dependence of the resistivity $\rho_a$ for different applied pressures. The arrows represent \tao at the considered pressure. Inset : Zoom of the resistivity curves around the \taot ordering transition. The transition is marked by the arrows. (b) Temperature dependence of the resistivity $\rho_c$ for different applied pressures. Inset : deduced pressure-temperature phase diagram of \reo. The closed symbols are the transition temperatures deduced from $\rho_a$ while the open symbols are deduced from $\rho_c$. sC = semiconducting state, SDW=spin density wave, SC=superconductivity, M1 and M2 are metallic phases}
\label{fig1}
\end{figure}

\begin{figure}
\centerline{\includegraphics[width=0.8\hsize]{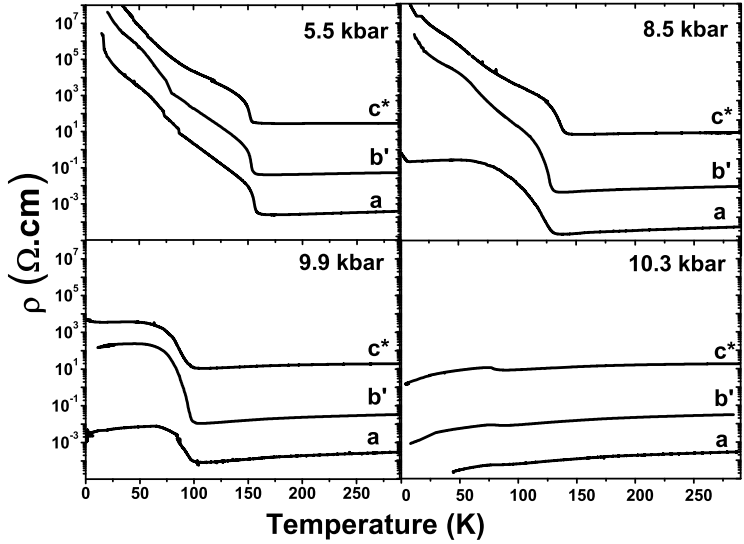}}
\caption{Resistivity of \reo along the \textit{a}, \textit{b'} and \textit{c*} axes at 5.5, 8.5, 9.9 and 10.3kbar.}
\label{fig2}
\end{figure}

Figure \ref{fig1} displays the temperature dependence of the resistivities $\rho_a$ and $\rho_c$ for different applied hydrostatic pressures, obtained upon cooling. The anion ordering transition \tao is determined using a standard criterion, i.e. the maximum of $d(ln\rho)/dT$. In agreement with previous observations \cite{Parkin81}, below 8kbar, this transition is weakly first order. For larger pressures, strong metastabilities and hysteresis up to 30K wide are visible which weaken only near the observed end point of the transition line. Similarly, \taot is clearly observed above 8 kbar as a cusp in the $\rho_a$ curve as shown in the inset of Fig.\ref{fig1}a and as a smooth change in the curvature of $\rho_c$. The deduced phase diagram, represented in the inset of Fig.\ref{fig1}b, is quite similar to the one already published using X-rays \cite{Moret86} or transport \cite{Parkin81} measurements. Figure \ref{fig2} presents, for four different applied pressures, the evolution of the resistivity along the three axes. At the lowest pressure, P=5.5kbar, in the homogeneous regime where the q$_2$ anion ordering extends over the whole sample, the anisotropy $\rho_c /\rho_a$ is decreasing when lowering the temperature. This observation is also valid for lower pressures and is compatible with the experimental data of Korin-Hamzic \textit{et al.} \cite{Korin03} obtained at ambient pressure where $\rho_c /\rho_a$ was experimentally shown to decrease by a factor of about 100 between 150K and 80K. As expected, $\rho_b$ follows an intermediate way between $\rho_a$ and $\rho_c$ in the continuity of the observations of Korin-Hamzic \textit{et al.} at ambient pressure. However, at P=8.5kbar, when both anion orderings may be present, the resistivities along the different axes strongly diverge below 80K as a metallic behavior is observed in $\rho_a$ while such a trend is observed neither in $\rho_b$ nor in $\rho_c$. For larger pressures, a metallic behavior at low temperature is recovered along the \textit{b'} direction but still delayed along the \textit{c*}-axis as shown at P=9.9kbar. Finally, at 10.3kbar, the q$_2$ anion ordering temperature is nearly washed out along all three directions. 

\begin{figure}
\centerline{\includegraphics[width=0.7\hsize]{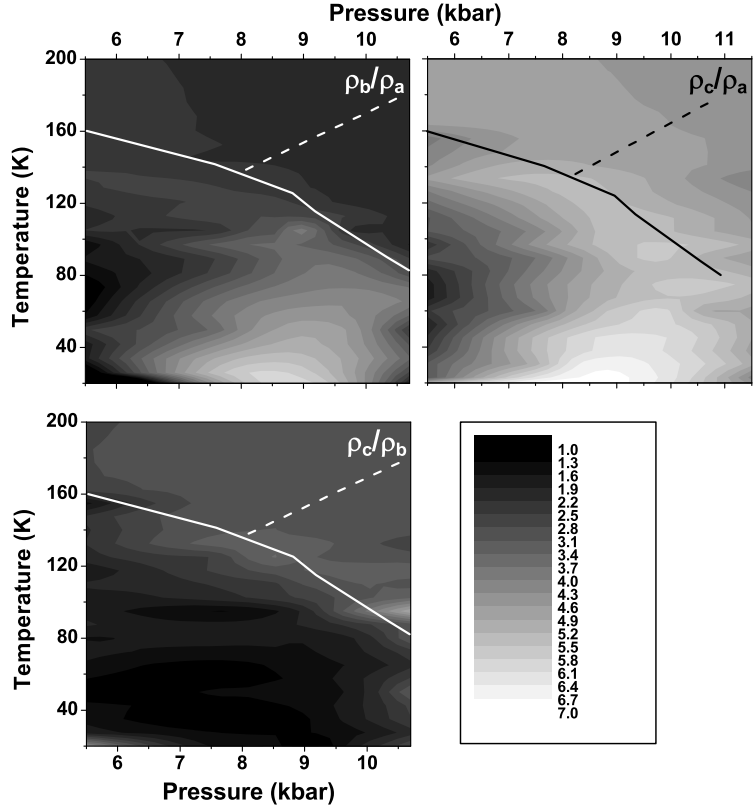}}
\caption{Surface plots of the logarithm of the anisotropies $\rho_b /\rho_a$, $\rho_c /\rho_a$ and $\rho_c /\rho_b$ of \reo as a function of temperature and pressure. The solid line represents schematically the first order phase transition line, \tao and the dashed line, the second order phase transition line, \taot.}
\label{fig3}
\end{figure}

Figure \ref{fig3} visualizes, on a surface plot, the evolution of the anisotropies $\rho_b /\rho_a$, $\rho_c /\rho_a$ and $\rho_c /\rho_b$ as a function of both pressure and temperature in the domain 20$<$T$<$200K. The gray scale is identical for the three plots and corresponds to a logarithmic scale with a variation of 7 orders of magnitude of the anisotropy from black to white. The strong variation of the anisotropy observed around the anion ordering temperature \tao might be attributed to the changes in the lattice parameters associated with the first order transition. Due to the logarithmic scale, the change of anisotropy observed upon crossing the second order phase transition at \taot is not visible even if it is effectively present. As shown in Fig.\ref{fig3}, on both $\rho_c /\rho_a$ and $\rho_b /\rho_a$ plots, large anisotropies have a tendency to penetrate deep below \tao at low temperatures. This is expected in charge transfer salts where decreasing the temperature is equivalent to increasing the pressure. The main point of this letter is the existence of a clear light spot at low temperatures on the $\rho_c /\rho_a$ plot centered on P=8.8kbar. This is the signature of the strong increase of the metallicity along the \textit{a}-axis  with respect to the \textit{c*}-axis at low temperatures. A similar observation can be deduced from the $\rho_b /\rho_a$ plot but is less marked. However, on the $\rho_c /\rho_b$ plot, no real structure appears demonstrating that \textit{c*}-axis and \textit{b'}-axis play more or less an equivalent role. On the other hand, \reo is a superconductor above 9.5kbar \cite{Parkin81} with a critical temperature, $T_c\leq$1.7K, even if superconductivity may be masked by the q$_2$ anion ordering \cite{Tomic84} and eventually by the spin density wave state. In this respect, the study of the superconducting transition gives strong indications on the continuity of the conduction along the different axes. As shown in Figure \ref{fig1}, below 10kbar, both spin density wave and superconducting states are successively stabilized suggesting an additional phase coexistence similar to the one observed in \pf\cite{Vuletic02}. These two instabilities originate from phase transitions of the metallic parts of the sample characterized by the q$_3$ order as shown by Tomi\`c \textit{et al.}\cite{Tomic89}. As a result, in a weak domain of pressure and temperature, we deal with a triple phase coexistence: a semiconducting state, a spin density wave state and a superconducting state are present. The superconducting transition is complete along \textit{a}-axis, \textit{i.e.} the resistance reaches the value $R=0$, only for pressures larger than 10kbar as shown in Fig.\ref{fig1}a. However, the zero resistance state is achieved along the \textit{c*}-axis only for P$\geq$10.6kbar as shown partially in Fig.\ref{fig1}b. In granular samples, a zero resistance state is just the signature of the existence of one conducting path  with Josephson coupled grains. On the other hand, the coincidence of the disappearance of the spin density wave state and the observation of the zero resistance state in the superconducting state suggests that the domains with q$_3$ order may already elongate throughout the sample at a lower pressure. However, our resistivity data are not enough to conclude on this point and we therefore define as 10kbar, the pressure above which q$_3$ order domains extend from one end of the sample to the other along \textit{a}-axis. 

The experimental image of the phase coexistence in \reo is then clear: below 10kbar, metallic droplets elongated along the \textit{a} axis are embedded in a semiconducting matrix characterized by the q$_2$ order. As pressure is further increased, conduction occurs through filaments made of several TMTSF chains. This filamentary texture prevails nearly up to the end of the coexistence regime detected as the end of the visibility of the \tao line. In cuprates, the observation of an electrical in-plane anisotropy up to 2.6 has been considered as a proof of one-dimensional self-organization of charge and the formation of stripes\cite{Ando02}. Here, in \reo, the anisotropies $\rho_b /\rho_a$ and $\rho_c /\rho_a$ show variations up to $10^3-10^4$ from their values in the high temperature metallic state which definitely establish the 1D character of conduction and the ultra-low probability of any wandering of the conducting paths along the transverse directions. In this sense, our data are much similar to the anisotropies, up to 100, observed in two dimensional electron gas (2DEG) in the quantum Hall regime at half filling factor of high Landau levels \cite{Lilly99}. However, self-organization of charge in \reo has a completely different origin from the one in the cuprates, manganites or 2DEG where phase separation occurs because of the competition of electronic instabilities. We are dealing, here, with a first order phase transition between two crystallographic orders which differ only by a factor of two in the lattice parameter along the chains (\textit{a}-axis). The observation of an endpoint of the \tao transition line is not physical as no second order phase transition from q$_2$ to q$_3$ orders is possible. This thermodynamic point of view should be temperated by kinetic arguments to modify the orientation of the anions from one position to the other. This effect of external parameters on anion ordering rate is also well known in \clo \cite{Joo05}. Due to this effect, the first order line seems to terminate at a critical point which was also obtained using a Helium gas cycling procedure by Tomi\`c \textit{et al.} \cite{Tomic89}. On the other hand, \reo is quite similar to $\beta$-(BEDT-TTF)$_2$I$_3$ where the incommensurate $\beta_L$ and ordered $\beta_H$ phases compete \cite{Kang87}. A model based on the existence of two energy minima of the local potential has successfully explained the main lines of its phase diagram \cite{Ravy88} highlighting a region in the phase diagram with a strong reentrance of the coexistence regime below the first order transition similar to our observations of Fig.\ref{fig3}. In order to understand the phase coexistence of the two anion ordering in \reo, an analogy with a ferromagnetic sample appears to be natural. Such Ising analogy has been already used in the understanding of anion orderings in the \tmtsfx family where the spins were taken as the orientation of the tetrahedral anions \cite{Ilakovac97,Pouget96}. Since the orientation of the tetrahedrons leads to a precise value of the \textit{a}-axis parameter, it is easier, here, to associate a complete unit cell parameter to the value of the spin. It appears then natural to map \reo into a 'rectangular' anisotropic ($J_x$,$J_y$) Ising lattice placed in an external magnetic field where a spin up(down) represents one cell with the parameter \textit{2a}(\textit{a} respectively). Since the anisotropy of metallic \reo in the (\textit{a}-\textit{b}) plane is about 10$^2$ and the lattice parameters verify \textit{a}$<$\textit{b}, we fix $J_x>>J_y$. The geometry of the domains for an anisotropic Ising lattice is well known to be essentially rectangular with an anisotropy given by $J_x /J_y$ \cite{Onsager44,Weng67,Zia82}. As a result, in \reo, the domains where the q$_3$ order is present are strongly elongated along the $\mathrm{TMTSF}$\, chains in agreement with our experimental observations. Therefore, we are dealing with a superstructure of macroscopic objects, droplets or filaments, coupled to their neighbors by the pressure dependent parameters, denoted $\tilde{t}_a$, $\tilde{t}_b$ and $\tilde{t}_c$ along the crystallographic axes \textit{a}, \textit{b} and \textit{c}, respectively. The properties of the conduction in the coexistence regime of \reo can now be well understood. At low temperatures, the effect of the pressure is similar to the one well known in homogeneous samples : at low pressures, $\tilde{t}_b$ and $\tilde{t}_c$ are negligible and conduction is essentially one-dimensional, the observed evolution of the anisotropy is associated to the expected increase of $\tilde{t}_a$ with pressure. For larger pressures, $\tilde{t}_b$ and $\tilde{t}_c$ might be considered and a transition toward a more two or three dimensional conduction is expected leading to a decrease of the anisotropy as observed above 8.8kbar (Fig.\ref{fig2} and Fig.\ref{fig3}). On the other hand, the effect of temperature at a fixed pressure is straightforward if we consider that hopping from one domain to another is thermally activated, washing out the anisotropy of conduction. Finally, we can also understand why, in our compound, the presented results have not been observed  using other techniques such as X-ray measurements \cite{Moret86} or other volume experiments. As filamentary conduction is now established in our compound, the low temperature metallic fraction, c(P), of the sample can be estimated from the resistivity at a temperature just above the spin density wave and superconducting transitions since the current flows only in the metallic regions at low temperatures. Then, the estimated fraction is the ratio of the \textit{a}-axis resistivity at 15K at P=10.6kbar, where \tao is no more visible, to the resistivity at 15K at the considered pressure. Using the data of Fig.\ref{fig1}a, we estimate c(8kbar)$\approx$10$^{-7}$, c(8.5kbar)$\approx$10$^{-5}$ and c(9.8kbar)$\approx$10$^{-3}$. The uncertainties may be larger than one order of magnitude but they demonstrate that electrical conductivity determination is certainly one of the most appropriate technique to study this regime.  

The study of the coexistence regime of \reo through transport anisotropy measurements has revealed its texture: metallic parts form droplets then filaments elongated along the most conducting direction upon increasing the pressure. The evolution of the anisotropy can be easily understood using a mapping into a rectangular Ising lattice. Such a determination in an organic conductor might open new experimental studies in strongly anisotropic materials at the metal-insulator boundary as transport measurements are very sensitive to few amounts of metallic regions. One important question is now to understand whether the presence of electronic instabilities may strongly affect the conclusions presented here such as in \pf for instance where a lamellar structure was suggested for the spin density wave / superconductivity phase coexistence. 

\acknowledgments
 The authors acknowledge D.J\'erome, C.Bourbonnais, P.Wzietek, J.P.Pouget and S.Tomi\'c for fruitful discussions. This work is supported by the European Community under Grant COMEPHS No. NMPT4-CT-2005-517039. One of us (C.C.) also acknowledges this grant for a partial financial support.

\end{document}